\documentclass[reprint, amsmath,amssymb,aps]{revtex4-2}

\usepackage{graphicx}
\usepackage{bm, slashed}
\usepackage{tikz} 
\usepackage{hyperref}
\usepackage{comment}
\usepackage{mathdots,bbold}
\usepackage{soul}
\usepackage[normalem]{ulem}
\setcitestyle{square,numbers}

\begin{document}

\title{Topological van Hove singularities at phase transitions in Weyl metals}

\author{Pierpaolo Fontana}
 \email{pfontana@sissa.it}
\affiliation{
 SISSA and INFN, Sezione di Trieste, Via Bonomea 265, I-34136 Trieste, Italy
}

\author{Michele Burrello}
\affiliation{
	Niels Bohr International Academy and Center for Quantum Devices, Niels Bohr Institute, University of Copenhagen, Universitetsparken 5, 2100 Copenhagen, Denmark
}
\author{Andrea Trombettoni}
\affiliation{%
 Department of Physics, University of Trieste, Strada Costiera 11, I-34151
Trieste, Italy
}
\affiliation{SISSA and INFN, Sezione di Trieste, Via Bonomea 265,
I-34136 Trieste, Italy}

\date{\today}

\begin{abstract}
We show that in three-dimensional (3D) topological metals a subset of the van Hove singularities of the density of states sits exactly at the transitions between topological and trivial gapless phases. We may refer to these as topological van Hove singularities. By investigating two minimal models, we show that they originate from energy saddle points located between Weyl points with opposite chiralities and we illustrate their topological nature through their magnetotransport properties in the ballistic regime. We exemplify the relation between van Hove singularities and topological phase transitions in Weyl systems by analyzing the 3D Hofstadter model, which offers a simple and interesting playground to consider different kinds of Weyl metals and to understand the features of their density of states. In this model, as a function of the magnetic flux, the occurrence of topological van Hove singularities can be explicitly checked. 
\end{abstract}

\maketitle

\section{Introduction}
Weyl semimetals \cite{wan2011,Burkov2011} are the simplest three-dimensional (3D) systems that combine a gapless spectrum with topological features. They are characterized by pairs of low-energy bulk excitations with linear dispersion, whose dynamics is described by the gapless Weyl equation. When their chemical potential matches the energy of the Weyl band-touching points, the Fermi surface shrinks to a discrete set of bulk points, and the low energy transport properties are dominated by the topological surface features.

By shifting the chemical potential away from the Weyl points, the Fermi surface becomes in general a collection of two-dimensional (2D) Fermi sheets \cite{Haldane2014,PRB_Okugawa_Murakami,Vanderbilt2015}  surrounding, in momentum space, each band-touching point. They are characterized by a non-zero Chern number, such that this state of matter is referred to as topological metal \cite{Burkov}. With suitable boundaries, topological metals are characterized by protected chiral surface states and the corresponding anomalous Hall conductivity. By varying the chemical potential further away, the Fermi sheets typically merge, recovering trivial non-topological metallic states. This process not only constitutes a Lifshitz transition \cite{Volovik2017}, due to the change of the topology of the Fermi surface, but it can also be seen as a topological phase transition (TPT) between two gapless states of matter, due to the change of the topological invariants associated with the Berry fluxes of the Fermi surface\footnote{To avoid any misunderstanding, we comment that very often the term TPT is used to denote a transition, as it occurs in 2D systems, where the gap closes and the topological invariants change. Here, we are considering 3D systems and the transition is between \textit{gapless} phases, with a change of the topological invariants.}. Therefore, not all the Lifshitz transitions are topological in this sense, and in the following we will refer to the non-topological Lifshitz transitions as \textit{standard} Lifshitz transitions. In this article, we show that the TPTs between Weyl and trivial metallic phases are accompanied by van Hove (VH) singularities, as in the case of standard Lifshitz transitions, and they are signaled, in general, by peculiar magnetotransport features resulting from the vanishing chiral anomaly (see also \cite{Louvet_2018}). 

The first realizations of Weyl semimetals in solid-state compounds \cite{felser2017,hasan2017,Armitage2018RMP} have been accompanied by successful implementations of their bosonic counterparts in photonic crystals \cite{lu2015,Yang2018} and ultracold quantum gases \cite{Wang2020}. Many artificial quantum materials have indeed been proposed for the simulation of topological semimetal, including, for instance, multiterminal Josephson junctions \cite{RiwarNatComm2016,RepinPRB2019,Repin2020,FatemiPRR2021}.

Proposals based on artificial quantum materials typically take advantage of the high level of tunability of their physical parameters. An important example is provided by the possibility of introducing sizable magnetic fluxes piercing the plaquettes of artificial crystals. For instance, moir\'e superlattices have been successfully used to explore features of the Hofstadter model \cite{Dean2013}. These developments inspired the study of Weyl semimetals with lattice models characterized by strong magnetic fluxes \cite{Affleck-Marston,Hasegawa,Laughlin_Zou,Lepori_2010,PRL_Ketterlee,PRB_Lepori_Fulga_Trombettoni_Burrello,Roy2016,shastri2017}. Even higher-dimensional extensions have been investigated, using synthetic dimensions or topological charge pumps \cite{GaneshanPRB2015,ZhangPRA2015,BlochTCPNature}.

The possibility of obtaining novel TPTs through the introduction of strong magnetic fluxes has been recently investigated in 2D Chern insulators \cite{Bernevig2020,Santos2020}. In \cite{Santos2020}, in particular, it was discussed how VH singularities in proximity to 2D Dirac points may cause TPTs with a large variation of the Chern number under small changes of the magnetic flux.

In 3D gapless systems, TPTs can be defined by changes of the topological invariants of the Fermi sheets, occurring at specific singular points. Additionally, Fermi surface singularities result in the presence of VH points, namely discontinuities in the energy derivative of the density of states (DOS). One can have a change in the topology of the Fermi surface, thus a Lifshitz transition \cite{Volovik2017}, without a TPT. Therefore, not all the VH points are associated with TPTs. We will show that the TPTs between 3D topological and trivial metals, occurring when the chemical potential is varied, are characterized and accompanied by the appearance of VH singularities in the DOS. 

As a case study, we will consider the 3D Hofstadter model, which offers a useful playground for modeling several topological metallic phases. Our choice is inspired by the recent progress in the engineering of quantum matter in artificial lattices with large effective magnetic fluxes, which encompasses many branches of many-body physics \cite{aidelsburger2018}, including driven ultracold atoms trapped in an optical lattice \cite{Aidelsburger2013,Miyake2013,goldman2014,Aidelsburger2015,weitenberg2021}, molecular nanostructures built with scanning tunneling microscopes \cite{kempkes2019,fremling2020}, moir\'e double-layer heterostructures \cite{Dean2013,Hunt2013}, and photonic crystals \cite{ozawa2019}.

The paper is organized as follows. In Sec. \ref{general concept} we introduce the main general concept by analyzing a continuum model showing ideal Weyl points. In Sec. \ref{chiralanomaly_signatures} we discuss some characteristic properties of the TPTs based on their chiral anomaly signatures; in particular, we investigate the ballistic magnetotransport features of two toy models in proximity to the TPTs. In Sec. \ref{hofstadter_model} we introduce the 3D Hofstadter model, and we present our result for generic magnetic fluxes. In Sec. \ref{conclusion} we summarize and present our conclusions. The Appendixes contain some details and computations for the Hofstadter model at different fluxes.

\section{General concept \label{general concept}}
\begin{figure}[t]
	\centering
	\includegraphics[width=1.05\linewidth]{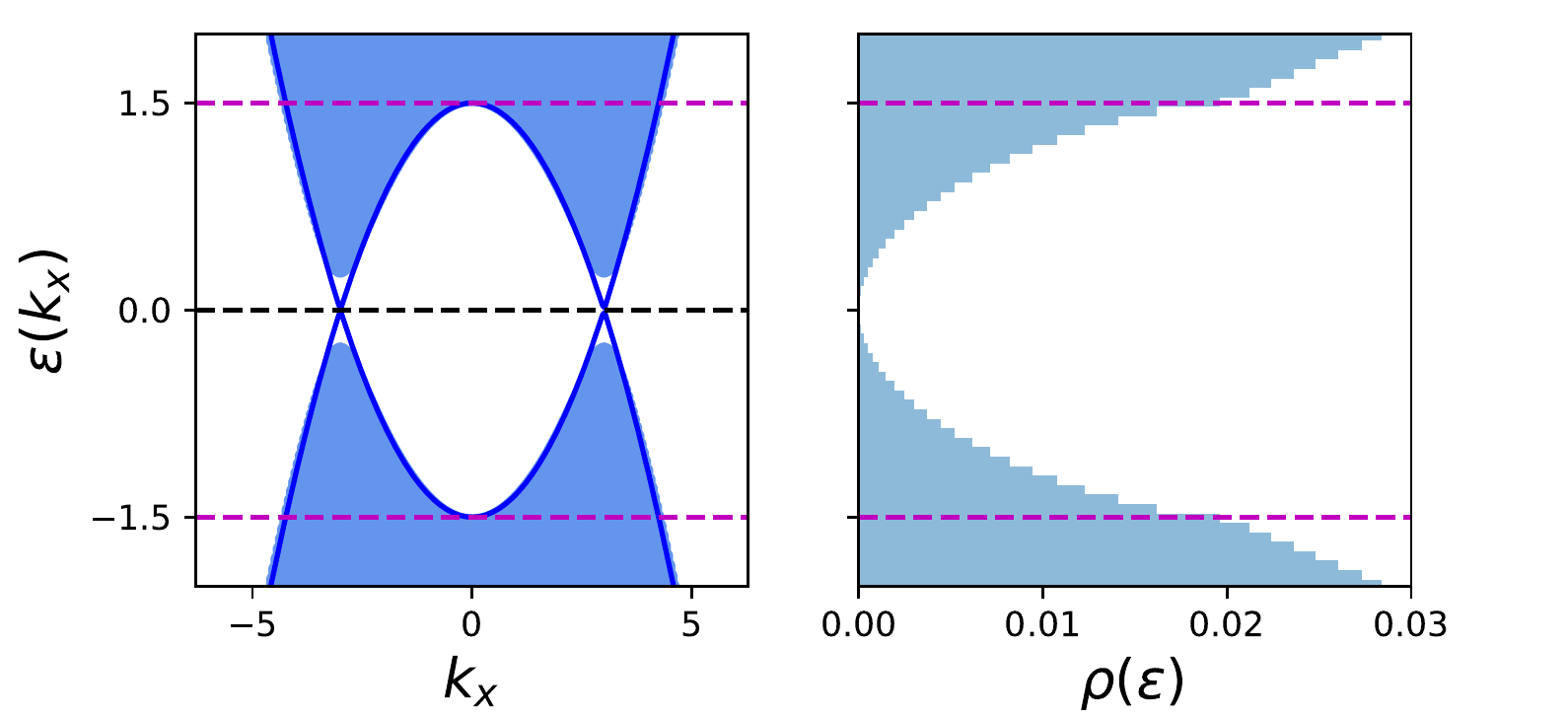}
	\caption{Dispersion relation of Eq. 
		\eqref{ham} for $k_z=0$, alongside the associated DOS. We highlight the Weyl points energy $\epsilon=0$ (black dashed line) and the VH singularities at $\epsilon=\pm E_s$ (magenta dashed lines). In this figure and in the following, energies are in units of $t$.}
	\label{toymodel_disprel_rotatedDOS}
\end{figure}

To begin our discussion we use the minimal two-band toy model \cite{PRB_Yang_Lu_Ran,PRB_Okugawa_Murakami,Lopez2018,Burrello_Guadagnini} defined by:
\begin{equation} \label{ham}
H({\bf k}) = \frac{{v}}{2k_0} \left(k_x^2-k_0^2\right)\sigma_x + vk_y \sigma_y + vk_z \sigma_z -\mu\, .
\end{equation} 
For $\mu=0$, the Hamiltonian \eqref{ham} describes a Weyl semimetal with two Weyl points of opposite topological charges located at ${\bf{k}}_W=\left(\pm k_0,0,0\right)$. Their linear dispersion is characterized by the same velocity $v$ in all directions and the corresponding DOS quadratically vanishes at zero energy. For small variations of $\mu$, the Fermi surface is composed by two (almost) spherical Fermi sheets of radius $|\mu/v|$ centered on each Weyl point. These Fermi sheets are characterized by Chern numbers $\pm 1$, matching the Weyl topological charges; thus the system is a topological metal \cite{Haldane2014,PRB_Okugawa_Murakami,Vanderbilt2015,Burkov}.
This topological phase survives until $|\mu| = vk_0/2$ (magenta lines in Fig. \ref{toymodel_disprel_rotatedDOS}). 
Here, the two Fermi sheets become connected in the point ${\bf{k}_s}=\left(0,0,0\right)$, with the result that the protected surface states completely overlap in momentum space with the zero-energy bulk states. For $|\mu| \ge vk_0/2$, there are paths within the Fermi surface connecting in momentum space the two Weyl points. $E_s \equiv vk_0/2$ is the minimum value of $|\mu|$ such that these paths open and, as a result, for $|\mu|=E_s$ the Fermi surface becomes a single connected surface with Chern number zero, since it encloses two Weyl points with opposite charges. This is indeed a TPT between a topological metal at $|\mu| < E_s$ and a trivial metal for $|\mu|>E_s$. 

In momentum space, ${\bf{k}_s}$ constitutes a saddle point for the energies of both bands, hence it gives rise to two VH singularities in the DOS $\rho(\epsilon)$ of the system at energies $\pm E_s$, which correspond to discontinuities in the derivative $d\rho/d\epsilon$, as shown in Fig. \ref{toymodel_disprel_rotatedDOS}. These VH singularities can be observed through bulk transport measurements or the investigation of optical properties via ARPES techniques \cite{PAL2018,Efremov2019,Lv2021RMP}.

The concurrence of the TPT between topological and trivial metals and the appearance of a VH singularity in the DOS of the system is a general feature of Weyl metals. To be more specific, let us consider two energy bands connected by two Weyl points with opposite topological charges. These Weyl points may have different energies and different velocities (we exclude, however, the case of strongly tilted type-II Weyl points \cite{Soluyanov2015,XuPRL2015}). For each path $\mathcal{K}$ between these two band-touching points in momentum space we can associate the differentiable function $\epsilon({\bf{k}})$ with $\bf{k} \in \mathcal{K}$, describing the lowest band energy along $\mathcal{K}$. (analogous results are obtained for the highest bands). The Weyl points are local maxima of the energy. Therefore, for each $\mathcal{K}$, there exists at least one global minimum $\epsilon({\bf{k_{\rm{min}}}})$. The TPT is located at $\mu=E_s$, where $E_s$ is the maximum of the energy minima $\epsilon({\bf{k_{\rm{min}}}})$ over the paths $\mathcal{K}$. This implies that ${\bf{\nabla}}_k \epsilon ({\bf k_s})=0$. Then, $E_s$ is a minimum of the energy along a path $\mathcal{K}$, but a maximum of the energy against variations of the path. Thus, in the most common scenario, it corresponds to a saddle point, and the vanishing of the velocity in this point causes a VH singularity at $\epsilon=E_s$ (other situations with stronger VH singularities are possible as well).

The previous argument can also be rephrased by applying the Morse theory \cite{Nash_Sen} to the differentiable function $\epsilon(\mathbf{k})$. In this framework, the critical points of $\epsilon(\mathbf{k})$, such that ${\bf{\nabla}}_k \epsilon ({\bf k})=0$, correspond to a change in the topology of the level sets of the function, i.e. the Fermi surfaces. In addition, the presence of the Weyl points ensure the change in the Chern number giving rise to a TPT. In realistic tight binding models, however, several additional VH singularities may appear that are not related to TPTs and that correspond to standard Lifshitz transitions. Therefore, in the following, we will distinguish \textit{topological} VH singularities from trivial VH singularities.

The topological VH singularities we discussed so far are based on a variation of the Chern number of the Fermi sheets by $\pm 1$. More complicated scenarios may be verified in systems presenting multi-Weyl points \cite{Bernevig_MultiWeyl_2012,Burrello_Lepori_DW,Huang1180,PhysRevB_Chen_2016,PhysRevB_Sbierski_2017}, or other kinds of band touching points with multiple topological charges \cite{Bernevig_UnconvWeyl_2016,Burrello_Fulga_TPC,PhysRevB_Fulga_Stern_2017}.

Let us consider, for instance, the case of double Weyl points with topological charge $\pm2$. In these models, characterized by the presence of additional symmetries, additional topological VH singularities can appear that are associated with the merging of two Fermi sheets with Chern numbers $\pm 2$. The set of topological phase transitions in these gapless systems, however, is richer: when considering models with multiple bands, further band touching points may cause topological phase transitions in which a Fermi sheet with Chern number $\pm 2$ split into two sheets with Chern number $\pm1$. These kinds of TPTs are not associated with VH singularities because they occur in correspondence with Weyl points connecting different bands, rather than saddle points appearing in a single band.

\begin{figure*}[t!]
	\centering
	\includegraphics[width=0.45\linewidth]{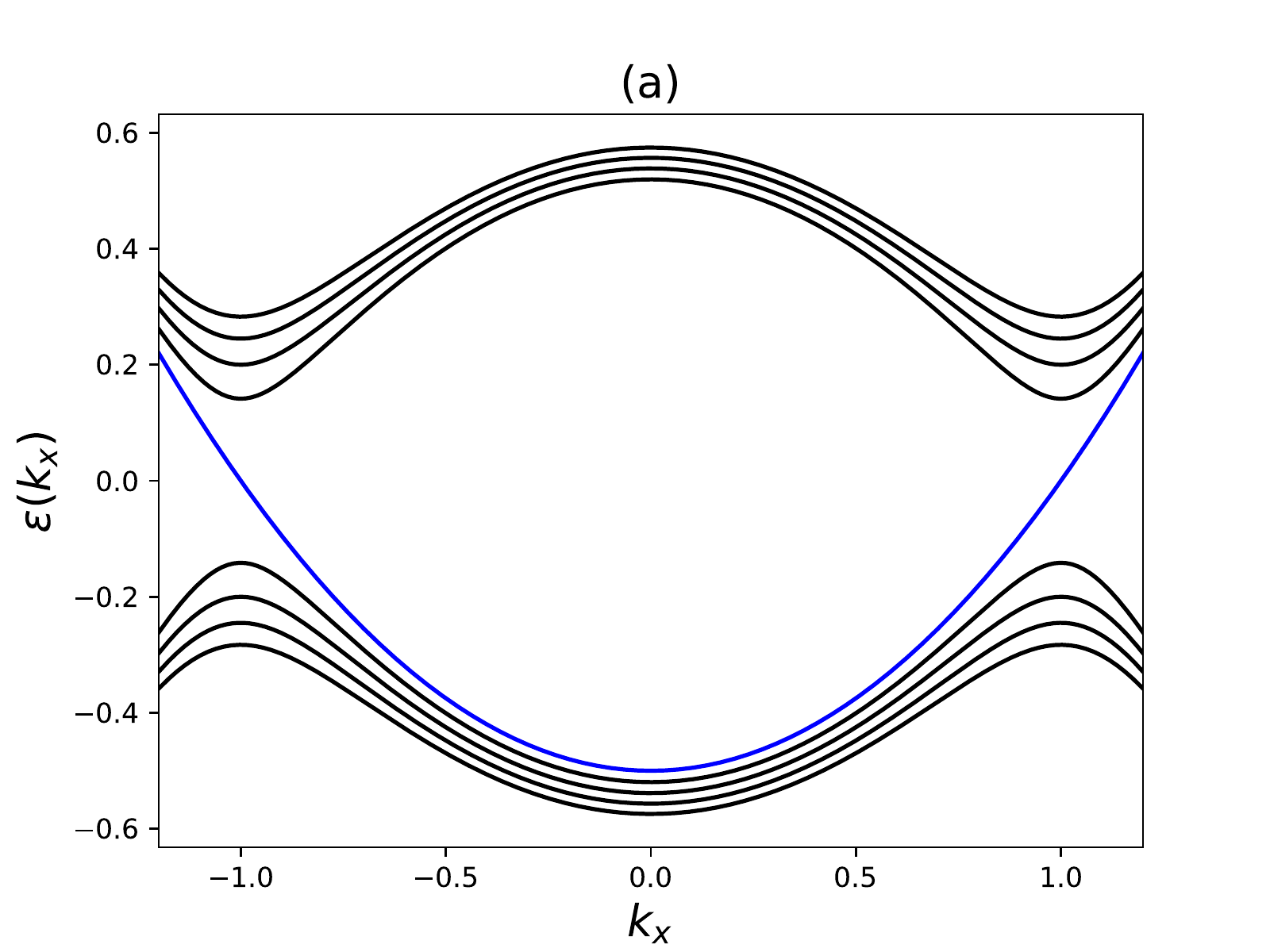}
	\hspace{0.3cm}
	\includegraphics[width=0.45\linewidth]{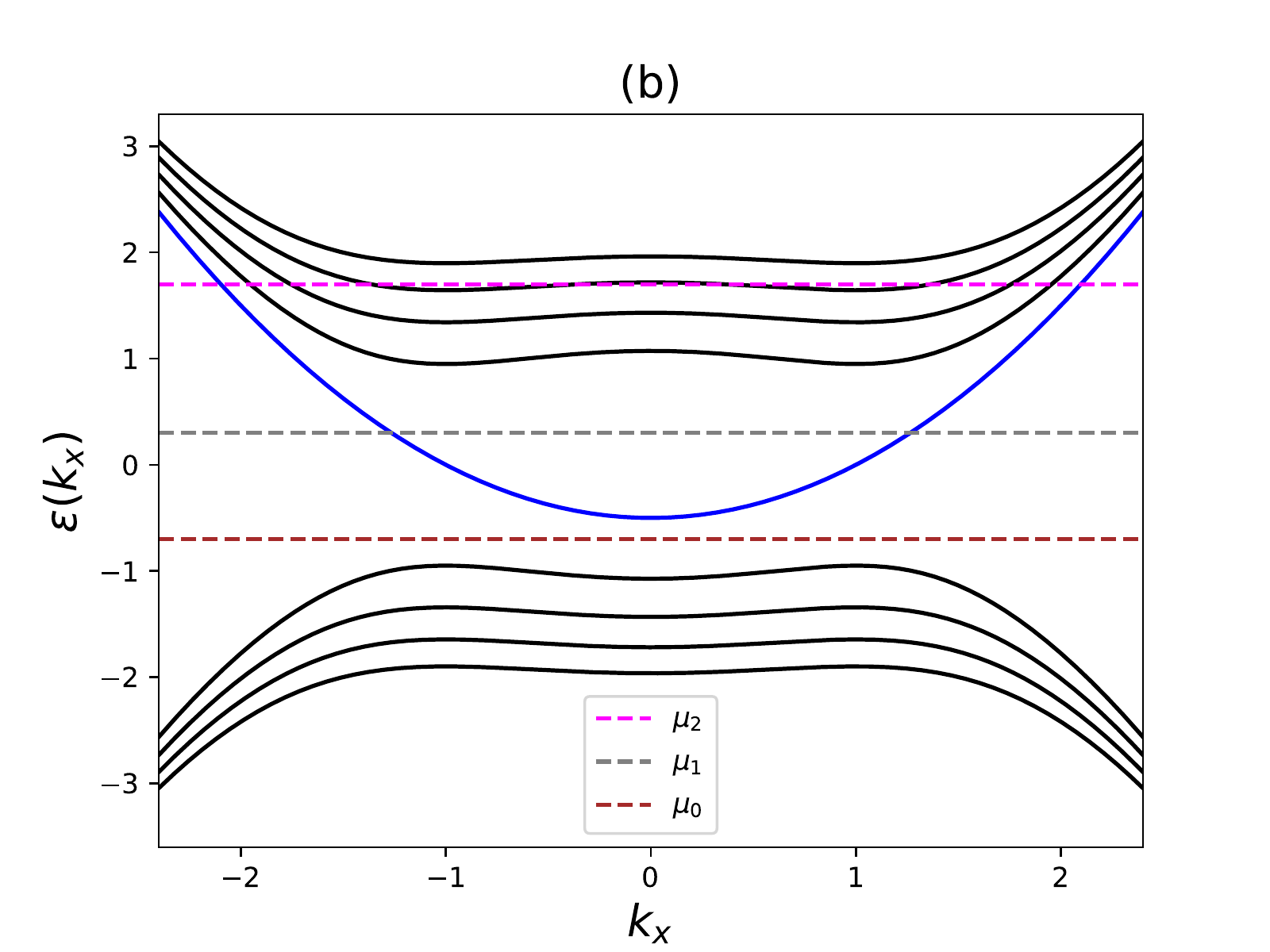}
	\caption{Landau levels $\epsilon_n(\mathbf{k})$ at fixed $k_y=k_z=0$, up to $n=5$, in the case of negative magnetic field in the two regimes $|B|<B_c$ (a), and $|B|>B_c$ (b). The chiral Landau level $(n=0)$ is plotted in blue. For large magnetic fields we distinguish the three regimes discussed in the main text: $\mu_0\in(-\sqrt{|B|},E_s)$ (brown dashed line), $\mu_1\in[E_s,\sqrt{|B|})$ (gray dashed line), and $\mu_2\ge\sqrt{|B|}$ (magenta dashed line). }
	\label{landaulevels_largefield}
\end{figure*}

\section{Signatures of the chiral anomaly in the topological phases\label{chiralanomaly_signatures}}

Across a TPT, the topological invariants associated with the Fermi sheets change: in a topological metallic phase, the system presents disjoint Fermi sheets with different Chern numbers, which give rise to signatures associated with the onset of chiral anomaly \cite{Burkov}. Chiral anomaly is indeed one of the main distinguishing features of the topological phases and in the following we investigate the related response of a Weyl system upon the introduction of parallel magnetic and electric fields, when the chemical potential varies across a TPT.

We refer in particular to the toy model in Eq. \eqref{ham}, and we consider the application of a magnetic field along the direction of the Weyl points, i.e. $\mathbf{B}=B\hat{x}$. The dispersion of the corresponding Landau levels $\varepsilon^{\pm}_n(k_x)$ is easily calculated as a function of ${B}$ \cite{Louvet_2018}:

\begin{equation} \label{landaun}
\varepsilon^{\pm}_n(\mathbf{k})=\pm\sqrt{|B|n+\frac{v^2(k_x^2-k_0^2)^2}{4k_0^2}}, 
\end{equation}
\begin{equation} \label{landau0}
\varepsilon_0(\mathbf{k})=-v\;\text{sgn}(B)\frac{k_x^2-k_0^2}{2k_0}.
\end{equation}

The Landau level at $n=0$ displays an opposite chiral behavior in proximity to the two Weyl points at $k_x = \pm k_0$. This chirality is exchanged when changing the sign of the magnetic field $B$, as a result of the different wavefunction of the $n=0$ Landau level. Hereafter we fix $B<0$, but the following considerations hold for $B>0$ and opposite energies. At fixed magnetic field, we can distinguish different regimes of the system depending on $\mu$.

For weak magnetic fields $|B|<B_c=E_s^2$, the system is gapless for any $\mu$, and indeed in Fig. \ref{landaulevels_largefield} (a) there is at least one partially filled Landau level for any chemical potential.
In this case, for $\mu<-E_s$ only non-chiral Landau levels contribute to the zero-temperature conductance of the system, while for $\mu>-E_s$ also the chiral $n=0$ Landau level conducts. Based on the construction by Nielsen and Ninomiya \cite{NIELSEN_CA}, we conclude that the conservation of chirality of the bulk states is broken for $\mu>-E_s$.

At stronger fields, $|B|>B_c$ (the so-called quantum limit), the situation is qualitatively different and we can identify four regimes: for $\mu\ge\sqrt{|B|}$ [see $\mu_2$ in Fig. \ref{landaulevels_largefield} (b)] all the Landau levels $\varepsilon^+_n(\mathbf{k})$ contribute to the zero-temperature conductance; for $\mu\in[-E_s,\sqrt{|B|})$ [see $\mu_1$ in Fig. \ref{landaulevels_largefield} (b)] only the chiral Landau level contributes to the bulk conductance; for $\mu<-E_s$ [see $\mu_0$ in Fig. \ref{landaulevels_largefield} (b)] an insulating phase appears; for other values of $\mu$ only the non-chiral Landau levels $\varepsilon^-_n(\mathbf{k})$ conduct.

In summary, for the trivial phase at $|\mu|>E_s$ there exists a critical value of the magnetic field amplitude $B_c$ above which an insulating phase appears. Within the topological phase at $|\mu|<E_s$, instead, the chiral Landau levels always contributes to the magnetoelectric transport, and, in particular, the system displays chiral anomaly. Another consequence of chiral anomaly is that in the topological phase for $|\mu|<E_s$, the system necessarily displays chiral Fermi arcs localized on any surface parallel to $\hat{x}$. In the trivial phase $|\mu|>E_s$, the Fermi arcs may instead vanish depending on the surface properties \cite{Burrello_Guadagnini}. This implies that the topological phase necessarily displays anomalous Hall conductivity, which, instead, may vanish in the trivial phase.

The above argument relies on the assumption of an ideal Weyl semimetal with a single pair of Weyl points. In the following, we extend it to a more general class of Weyl systems, characterized by multiple pairs of Weyl points. 

We focus in particular on systems displaying a collection of well-separated Weyl dipoles, each constituted by two Weyl points with opposite charges displaced by a small momentum distance. Our analysis of the Landau levels in Eqs. \eqref{landaun},\eqref{landau0} can indeed be generalized to  Hamiltonians of the form  $H (\mathbf{k}) = \sum_{j=x,y,z} g_j(k_j)\sigma_j -\mu$, and, to present a concrete example, we consider the following model of spin 1/2 fermions on a cubic lattice:
\begin{multline}
H_{\rm lat} = -\frac{v}{2\sin k_0} \sum_{{\bf r}} \left(c^{\dag}_{{\bf r} + \hat{x}} \sigma_x c_{\bf r} + {\rm H. c.}\right) + \\v \cot k_0\sum_{{\bf r}}c^{\dag}_{{\bf r}} \sigma_x c_{\bf r} 
- \mu \sum_{{\bf r}}c^{\dag}_{{\bf r}}  c_{\bf r}  + \\
\frac{v}{2} \sum_{j=y,z} \sum_{{\bf r}} \left( ie^{i \theta_j({\bf r})} c^{\dag}_{{\bf r} + \hat{j}} \sigma_j c_{\bf r}+ {\rm H. c.}\right)  \, .
\label{latHam}
\end{multline}
Here and in the following we set the lattice spacing to unit value. When the phases $\theta_j$ vanish, no magnetic field is present and the model displays four Weyl dipoles at $\mu=0$, each oriented along $k_x$ and with charges separated by $2k_0$, which we assume to be much smaller than $\pi$. In this case, $H_{\rm lat}$ describes a system that replicates four times the toy model in Eq. \eqref{ham}, with the four Weyl dipoles displaced by $\pi$ in the $k_y$ and $k_z$ directions of the BZ, and characterized by alternating orientations. The topological VH singularities lie at energy $E_s=v \tan (k_0/2)$.

\begin{figure}[h]
	\centering
	\includegraphics[width=1.0\linewidth]{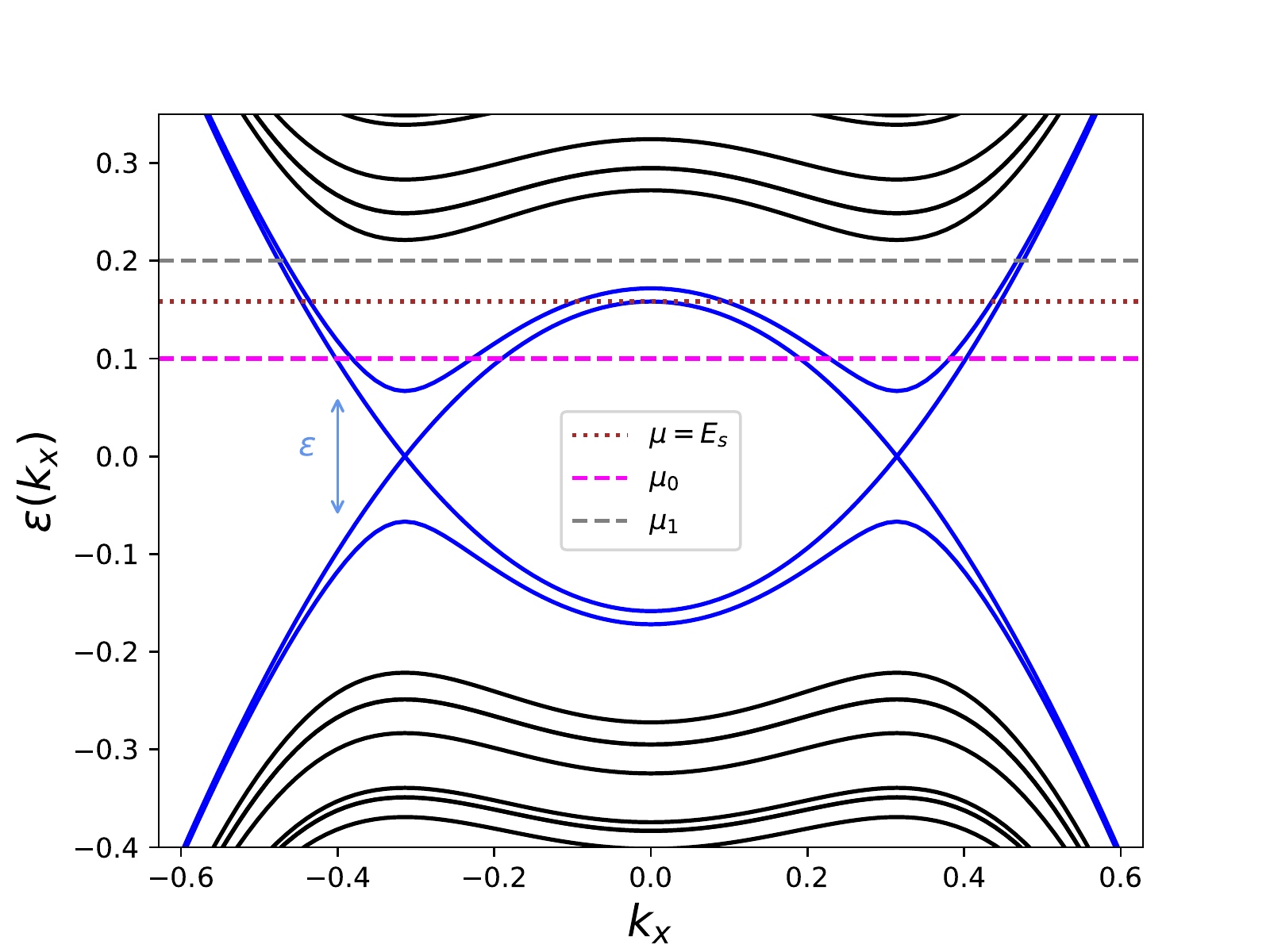}
	\caption{Landau levels $\varepsilon_n(\mathbf{k})$ of the model in Eq. \eqref{harper} at fixed $k_z=\pi/2$, in the case of positive magnetic field $B\gtrsim B_c$. The chiral Landau levels ($n=0$) are plotted in blue. We show two of the four regimes discussed in the main text: $\mu_0\in[\epsilon,E_s]$ (magenta dashed line), $\mu_1\in[E_s,\varepsilon_{1,\text{min}}]$ (grey dashed line). The saddle point energy $E_s$ (dotted brown line) is reported too.}
	\label{landaulevels_weyldipoles}
\end{figure}

The Peierls phases $\theta_j$ introduce a magnetic field in the system, and, analogously to the previuous example, we consider a magnetic field $\mathbf{B}=B\hat{x}$ parallel to the Weyl dipoles. We adopt a Landau gauge and we set $\theta_y({\bf r}) = 0$ and $\theta_z({\bf r})=By$. In this case, $k_x,\;k_z$ are conserved momenta, and the problem is effectively reduced to a collection of 1D systems:
\begin{multline} \label{harper}
	H(k_x,k_z) = \frac{v}{2} \sum_y\left( ic^{\dag}_{y+1} \sigma_y c_{y}+ {\rm H. c.}\right)-\\
	v\sum_{y} c^\dag_{y}\left[\frac{\left( \cos k_x - \cos k_0 \right)}{\sin k_0}\sigma_x 
	+ \sin\left(k_z+By\right)\sigma_z +\mu\right] c_{y}. 
\end{multline}
This choice of the magnetic field gives rise to a set of Landau levels for each Weyl dipole, analogous to the ones discussed for the ideal Weyl model. In particular, we expect two chiral Landau levels approaching zero energy and $k_x=\pm k_0$ for each chirality. Based on commensurability effects between $B$ and the Weyl point momentum distance, however, these chiral Landau levels may display avoided crossings, determined by the non-linear momentum dependence along $k_y$ and $k_z$ of the original lattice model (see \cite{Chan2017,Kim2017,Saykin2018}). Depending on the value of $B$, these avoided crossings may affect all the chiral Landau levels, or only a subset (as in the case of Fig. \ref{landaulevels_weyldipoles}), and they introduce a further energy scale in the problem, which we label with $\epsilon$.  This splitting of the chiral Landau levels can be considerable (see Fig. \ref{landaulevels_weyldipoles}) and was discussed in \cite{Chan2017,Saykin2018} for the ideal Weyl system \eqref{ham} with a magnetic field orthogonal to the Weyl point separation. However, an estimate of this energy scale in our model is a non-trivial problem, related to the solution of the generalized Harper model obtained by setting $k_x=\pm k_0$ in Eq. \eqref{harper}.  Intuitively, we expect that $\epsilon$ grows with $B$, and with the ratio $2k_0/\pi$ between the separation in momentum space of the Weyl points in the same dipole and the momentum distance of different dipoles, but given its non-monotonic behavior in $B$, the ballistic transport properties of the system \eqref{harper} are difficult to predict around zero energy.

Another effect of the non-linear perturbations of the dispersion of the cones, is that the critical field $B_c$ for the onset of the quantum limit decreases below $E_s^2$. For a weak splitting of the chiral Landau levels, at fixed $B\gtrsim B_c$, we identify four main regimes (see Fig. \ref{landaulevels_weyldipoles}): (i) for $\mu \lesssim \epsilon$, the conductance depends on the detail of the splitting of the chiral Landau levels, and, for specific values of $B$ and system sizes, insulating phases may appear (consistently with the analysis in \cite{Chan2017,Kim2017,Saykin2018}). (ii) For $\epsilon \lesssim |\mu| \lesssim E_s$, instead, the chiral Landau levels stemming from all the monopoles contribute to the bulk conductance $G$ (see $\mu_0$ in Fig. \ref{landaulevels_weyldipoles}); thus, in the limit of ballistic transport, $G = 4\left(e^2/h\right)\left(L^2/2\pi l_B^2\right)$, where $L^2$ is the area of the section of the system orthogonal to $\bf B$, and $l_B$ is the magnetic length of the system. (iii) For $|\mu|\in(E_s,\varepsilon_{1,{\rm min}}]$, where $\varepsilon_{1,{\rm min}}$ is the minimum of the energy of the non-chiral Landau levels, only the chiral Landau levels with a given chirality contribute to the transport (see $\mu_1$ in Fig. \ref{landaulevels_weyldipoles}), therefore $G = 2\left(e^2/h\right)\left(L^2/2\pi l_B^2\right)$. (iv) Finally, for $|\mu|>\varepsilon_{1,{\rm min}}$ also the Landau levels with $n \neq 0$ contribute to the conductance, increasing it at least by $\Delta G = 4\left(e^2/h\right)\left(L^2/2\pi l_B^2\right)$. In conclusion, at $|\mu|=E_s$ we expect a considerable variation in the number of channels contributing to the magnetotransport properties of the system in the large magnetic field regime; this gives rise to a peculiar discontinuity in the conductivity which drops from $G = 4\left(e^2/h\right)\left(L^2/2\pi l_B^2\right)$ on the topological side to $G = 2\left(e^2/h\right)\left(L^2/2\pi l_B^2\right)$ on the trivial side.

Importantly, the aforementioned transport features do not appear, in general, if the system undergoes a standard Lifshitz transition.

\section{The 3D Hofstadter model\label{hofstadter_model}}

\begin{figure*}[t!]
	\centering
	\includegraphics[width=0.4\linewidth]{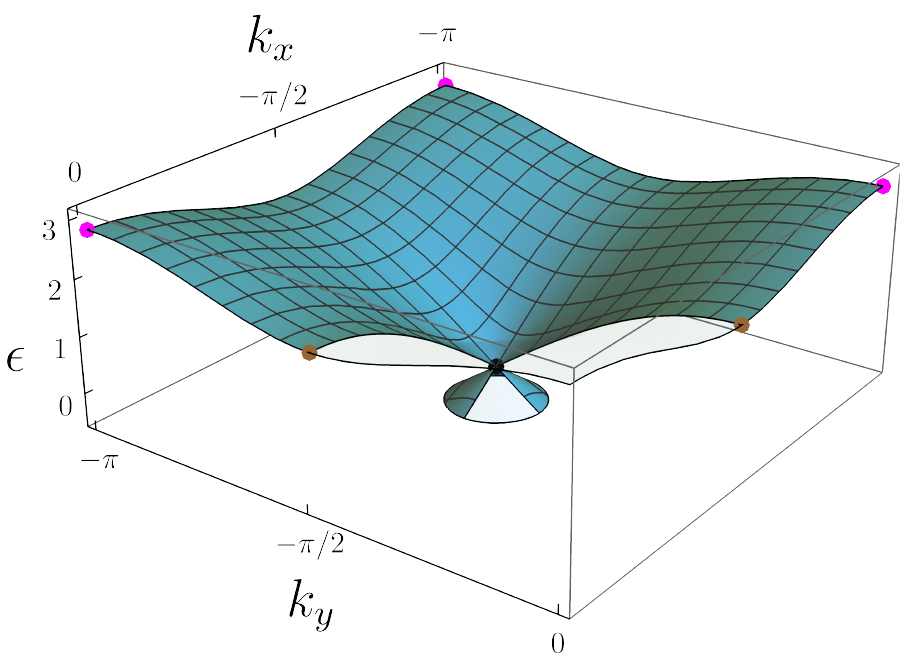}
	\hspace{0.3cm}
	\includegraphics[width=0.4\textwidth]{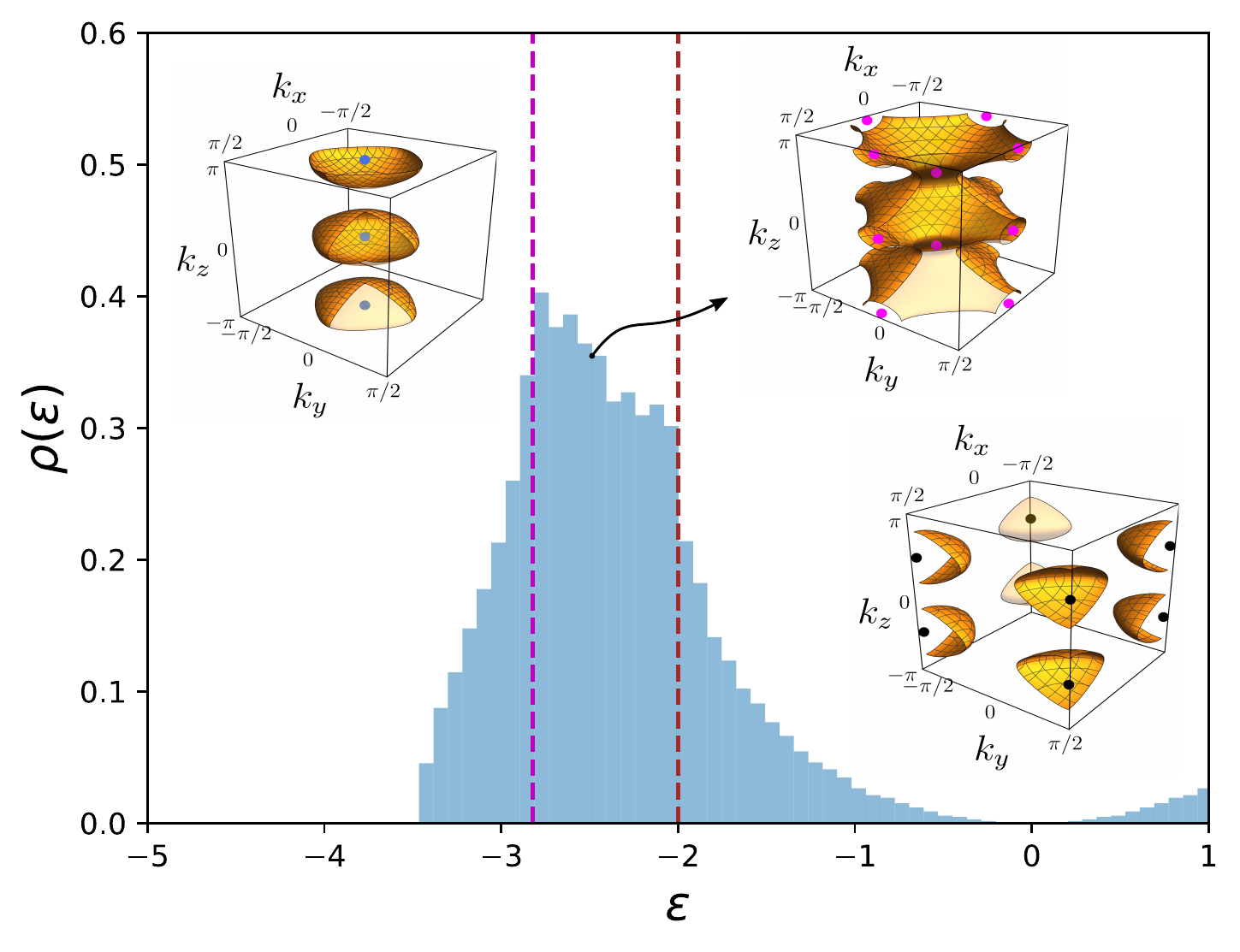}
	\caption{Left panel: second band in Eq. \eqref{2n_disprel} for $k_z=\pi/2$ centered around the point $\mathbf{k}_{\mathbf{W}}^{-,-}$, together with the topological (brown dots) and trivial (magenta dots) VH singularities. Right panel: Hofstadter model DOS for $n=2$ and $\epsilon\leq1$. We highlight the topological (brown dashed lines) and trivial (magenta dashed lines) VH singularities. Insets: Fermi surfaces in the MBZ for the different regions of $\mu$ explored in the text, highlighting the main points in momentum space.}
	\label{bands_DOS_FS_2n_plots}
\end{figure*}

To investigate how trivial and topological metallic phases alternate in lattice models and the corresponding patterns of VH singularities, we consider, as a case study, the 3D Hofstadter model. Its Hamiltonian on a cubic lattice reads  
\begin{equation} \label{hof}
H = -  t \, 
\sum_{{\bf r} \, , \, \hat{j}}   c^{\dagger}_{{\bf r} + \hat{j}} 
e^{i \theta_{j}\left({\bf r}\right)} 
c_{{\bf r} } + \ \mathrm{H.c.} \, 
\end{equation} 
Here the Peierls phases $\theta_j$ define the magnetic fluxes across all plaquettes. We consider the case of commensurate magnetic fluxes $\Phi=2\pi/n$ in all plaquettes 
($n \in \mathbb{N}$), corresponding to a magnetic field ${\bf B}= \Phi \left(1,1,1\right)$ in units of the magnetic flux quantum. 

The 2D Hofstadter model \cite{Hofstadter} is a paradigmatic example for the study of Chern insulators and the physics of the quantum Hall effect. Its spectrum as a function of the magnetic flux is the celebrated fractal Hofstadter butterfly. The 3D cubic lattice version of the model presents a complex and fractal spectrum as well \cite{koshino2001,koshino2002}. For $n=2$, with $\pi$ flux, it realizes a Weyl semimetal at half filling \cite{Affleck-Marston,Laughlin_Zou,Lepori_2010,PRL_Ketterlee} which, in particular, does not break the physical time-reversal and space inversion symmetries \cite{PRB_Lepori_Fulga_Trombettoni_Burrello}. For $n\neq2$, time-reversal symmetry is broken and the system displays $n$ connected energy bands. The spectrum is in general symmetric around zero energy due to the chiral sublattice symmetry. Therefore, we can restrict our analysis to negative energies.

\subsection{Varying the magnetic flux}

To study the model (\ref{hof}) it is convenient to choose the Hasegawa gauge \cite{Hasegawa} $\mathbf{A}(\mathbf{r})=\Phi(0,x-y,y-x)$ for the definition of the $\theta$ phases \cite{Burrello-Trombettoni1}. In the following, the behavior of the DOS, and the properties and location of the VH singularities, are studied for different values of $n\ge2$.

The Weyl semimetal appearing at $n=2$ is characterized by the band dispersion (see Appendix \ref{diagonalization_momentumspace}):
\begin{equation}\
\epsilon(\mathbf{k})=\pm 2t\sqrt{\cos^2k_x+\cos^2k_y+\cos^2k_z}.
\label{2n_disprel}
\end{equation}
It presents four inequivalent Weyl nodes in the magnetic Brillouin zone (MBZ), at ${\bf k_W}^{\pm,\pm}= \left(\pm \pi/2,\pm \pi/2, \pi/2\right)$. When $|\mu| < E_s \equiv 2t$, the Fermi surface is composed of four sheets with Chern number $\pm 1$. As expected, at $\mu=\pm 2t$ the system undergoes a TPT, and the Fermi sheets merge in a single surface with vanishing Chern number. This phase transition is accompanied by a topological VH singularity, generated by the  saddle points in ${\bf k_s} = \left(\pm \pi/2,\pm \pi/2, 0\right)$ (and the corresponding permutations within the MBZ; see the brown dots in Fig. \ref{bands_DOS_FS_2n_plots}). In the lowest band, the energy Hessian $\partial_{k_i} \partial_{k_j} \epsilon$ at the saddle points is diagonal with eigenvalues $2t\left\{1,-1,-1\right\}$, corresponding indeed to a minimum of the energy along the line joining two opposite Weyl points and to a local maximum in the orthogonal directions (the opposite happens in the highest band). 

By inspecting the DOS and the spectrum, we clearly see other singularities at $\epsilon = \pm 2\sqrt{2} t$ corresponding to additional saddle points at $(0,0, \pi/2)$ and the analogous points; see Fig. \ref{bands_DOS_FS_2n_plots}. These additional VH singularities can be understood by considering the behavior of the system for $-2\sqrt{3}t < \mu < -2\sqrt{2}t $, close to the minimum of the lowest band. In this regime, the Fermi surface is constituted by inequivalent disconnected sheets in the MBZ surrounding each band minimum (upper left inset in fig. \ref{bands_DOS_FS_2n_plots}). Each sheet has a vanishing Chern number. When $\mu = -2\sqrt{2}t$, these Fermi sheets merge in a single surface with zero Chern number (upper right inset in fig. \ref{bands_DOS_FS_2n_plots}), and the corresponding saddle points of the spectrum have a diagonal Hessian with eigenvalues $2t\left\{1,1,-1\right\}$. The topology of the Fermi surface changes at this energy, thus the system undergoes a Lifshitz phase transition between two topologically trivial metallic phases \cite{Volovik2017}. Therefore, we label the VH singularities at $\mu= \pm 2\sqrt{2} t$ as topologically trivial.

\begin{table*}[t!]
	\setlength{\tabcolsep}{10pt}
	\begin{tabular}{c|cccccccc}
		\hline 
		\hline
		$n$ &  \multicolumn{8}{c}{Energy (Weyl points $\to$ \textbf{boldface}; Topological VH singularities $\to$ \textcolor{red}{red})} \\ \hline
		2 & 
		\multicolumn{8}{c}{\hspace{0.3cm}
			\textcolor{red}{-2} \textbf{0}  \textcolor{red}{2}}\\
		3 & \multicolumn{8}{c}{ \textcolor{red}{-2.42}  $\mathbf{-\sqrt{3}}$  \textcolor{red}{-1.03}\hspace{1cm} \textcolor{red}{1.03}  $\mathbf{\sqrt{3}}$  \textcolor{red}{2.42}}\\
		4 & \multicolumn{8}{c}{\textcolor{red}{-2.70}  \textbf{-2.17}  \textcolor{red}{-1.63} \hspace{1cm} \textcolor{red}{-0.34}  \textbf{0}  \textcolor{red}{0.34}\hspace{1cm} \textcolor{red}{1.63}  \textbf{2.17}  \textcolor{red}{2.70}}\\
		5 & \multicolumn{8}{c}{\textcolor{red}{-2.71}  \textbf{-2.34} \textcolor{red}{-1.96} \hspace{5.5cm}
			\textcolor{red}{1.96}  \textbf{2.34} \textcolor{red}{2.71}}\\
		6 & \multicolumn{8}{c}{\textcolor{red}{-2.67}  \textbf{-2.43} \textcolor{red}{-2.18} \hspace{6.3cm} \textcolor{red}{2.18}  \textbf{2.43} \textcolor{red}{2.67}}\\
		7 & \multicolumn{8}{c}{\textcolor{red}{-2.60} \textbf{-2.50} \textcolor{red}{-2.40} 
			\hspace{7.1cm} \textcolor{red}{2.40}  \textbf{2.50} \textcolor{red}{2.60}}\\
		\hline
		\hline
	\end{tabular}
	\caption{Energies of the Weyl points and their associated topological VH singularities for $2\le n <8$.
	}
	\label{summary_hofst_results}
\end{table*}

Similar features characterize the Hofstadter model with $n=3$. In this case, the three bands of the model are separated by inequivalent pairs of Weyl points: within the MBZ there are two Weyl cones between the first two bands located at energy $\epsilon= -\sqrt{3}t$ \cite{Hasegawa}, and the symmetric cones between the second and third band at $\epsilon=\sqrt{3}t$. The system is thus in a Weyl semimetal phase when the chemical potential matches these energy levels, and the DOS vanishes quadratically at $\epsilon = \pm\sqrt{3}t$. By varying $\mu$ around the Weyl points, the system is in a metallic Weyl phase with disconnected Fermi sheets; the lowest and highest bands behave qualitatively as the bands in the $n=2$ model. The central band, instead, presents two TPTs from the topological phases at $|\mu| > t$ to a trivial phase for $|\mu|< t$. Besides these topological VH singularities, we can detect other VH singularities corresponding to local extrema of the three energy bands. At the level of the Fermi surface, these singularities signal the appearance of particle or hole pockets, thus of trivial Fermi sheets, and do not change the topological properties of the system.

Let now consider $n\ge3$. For $n$ smaller than a ``critical" value $n_c$, found to be $n_c=8$, the DOS of the system shows isolated zeros at $\epsilon=\mp \epsilon_{w,n}$, corresponding to Weyl points that separate the lowest (and the highest) energy bands from the others. This result for $n_c$ is consistent with the observation by Hasegawa \cite{Hasegawa} that, for $\Phi \lesssim 4\pi/31$, all the bands in the model overlap in energy. For $n<8$, the filling at the Weyl point with negative energy is thus $\nu=\frac{1}{n}$, generalizing the well-known result for $n=2$.

Around the energies $\epsilon_{w,n}$, the band structure is qualitatively similar to the $n=2$ case: there are two energy thresholds $E_{s,n}^{\mp}$, corresponding to saddle points, which determine topological VH singularities, thus the system is in a topological metal phase for $E_{s,n}^- <|\mu| < E_{s,n}^+$. Additional trivial VH singularities appear at different energies, signaling trivial Lifshitz transitions.

The case $n=4$ stands on its own, since a zero of the DOS is found also at $\epsilon=0$, corresponding to two Weyl cones.
These points overlap in energy with additional local extrema of the central bands, which are quadratically tangent to $\epsilon=0$. As a consequence, in this case the DOS vanishes as $\rho(\epsilon)\sim\sqrt{|\epsilon|}$, and not quadratically, at $\epsilon=0$, and the topological saddle points are doubled accordingly (see Appendix \ref{W_tildeWpoints_4n}). In particular, for $\mu\simeq0$, the Fermi surface is made by four disconnected sheets: two of them enclose the bands' stationary points, with Chern numbers equal to zero, while the remaining enclose the Weyl points, with Chern number $\pm1$.

Concerning smaller fluxes, i.e. $n \ge n_c$, $\rho(\epsilon)$ does not display any zero. As a consequence, there are ranges of $\mu$ for which the system is in a multiband metallic state, whose Fermi surface contains sheets generated from consecutive bands. 

Nonetheless, the system presents Weyl points at the energy $\epsilon=E_w$ between the first and the second band (and possibly between higher bands), whose presence  is ``hidden" by the band overlap. In this case, an additional VH singularity appears, which is associated with the minimum of the second band, with energy $E_m<E_w$ (see Appendix \ref{HHmodel_VH}). This singularity signals a Lifshitz non-TPT between two topological metals, i.e. a single band metal for $\mu<E_m$ and a multiband metal for $\mu>E_m$, which corresponds to the opening of an electron pocket. The topological VH singularity is located at the lower energy $E_s^- < E_m$. Similar hidden Weyl points appear also between the intermediate bands for $n=5,6,7$.

Our results on the Weyl points and their corresponding topological VH singularities for the Hofstadter model with $n<8$ are summarized in the Table \ref{summary_hofst_results}.

To conclude this Section, we comment about the characterization of the TPTs in the 3D Hofstadter model using the chiral anomaly ballistic transport signatures discussed above. The study of the magnetotransport in this model is a non-trivial task due to the commensurability effects induced by the presence of the magnetic field \cite{Roy2016}. 
On general ground, the Landau levels analysis presented in Sec. \ref{chiralanomaly_signatures} can be applied to the 3D Hofstadter model when considering small variations of the magnetic fluxes around a commensurate value in finite-size systems. In particular, a small flux perturbation of strength $|\delta\phi|=2\pi/q$ around the value $\phi=2\pi/n$ (with $q\gg n$) introduces a volume scale dependent on $q$ below which the Nielsen-Ninomiya argument for the appearence of the chiral anomaly is expected to hold, in analogy to similar results for the DOS obtained in \cite{Roy2016,PRB_Lepori_Fulga_Trombettoni_Burrello}. A full analysis of the dependence on the magnetic perturbation of the Weyl points in the 3D Hofstadter model and their magnetotransport properties is a subject certainly deserving of further study. 

\section{Conclusions\label{conclusion}}
We showed that the topological phase transitions (TPTs) in Weyl metals are signaled by the appearance of Van Hove (VH) singularities. As a function of the chemical potential, these transitions occur between a trivial phase, with a connected Fermi surface, and a topological phase, displaying disconnected Fermi sheets with non-zero Chern numbers. 
The related topological VH singularities manifest themselves as cusps in the DOS and are caused by the saddle points of the momentum space paths joining Weyl points with opposite chiralities. Their transport signatures may be enhanced for models displaying higher-order saddle points, whereas, in general, we expect them to vanish for strongly tilted (type-II) Weyl semimetals \cite{Soluyanov2015,XuPRL2015}.

To characterize these TPTs between different metallic states, we investigated some of the effects of the chiral anomaly and the Landau levels structure of these systems. Indeed, the chiral anomaly  gives rise to peculiar behaviors in the magnetotransport of Weyl semimetals \cite{Xiong_CA_2015,Huang_CA_2015,Hui_CA_2016,Niemann_CA_2017} and it allows us to distinguish topological and trivial phases. In particular, we studied the ballistic bulk magnetoconductance of relevant Weyl metal toy models. We verified that for sufficiently high magnetic fields with suitable orientation, the conductance of these systems displays a characteristic reduction by varying the chemical potential across a topological VH singularity from the topological to the trivial phase. In the case of ideal Weyl systems in this extreme quantum limit, insulating phases may appear in the trivial phase under strong magnetic fields \cite{Chan2017,Kim2017,Saykin2018,Louvet_2018}; for models with multiple pairs of Weyl points instead, the most common scenario corresponds to halving of the conductance. This discontinuity of the magnetoconductance is typical of the topological VH transitions, and it does not appear for standard Lifshitz transitions.

As an illustrative example of the onset of topological VH singularities, interesting \textit{per se}, we investigated the Hofstadter model on a 3D cubic lattice as a function of the magnetic flux $\Phi$ of the form $\Phi=2\pi/n$, which hosts several trivial and topological gapless phases when $n$ is varied, and is relevant for the study of novel superconducting materials \cite{Ran2019,Park2020}. For $n<8$, its lowest bands do not overlap, and we identified the TPTs by inspecting directly the singular points of the DOS. For $n \ge 8$ the bands overlap, and Lifshitz transitions between single and multiband metallic phases can be identified. Our analysis opens the possibility of definining a 3D analog of the Hofstadter butterfly by distinguishing it as a function of the flux and the chemical potential trivial and topological gapless phases. 

Analogous features can be investigated for generic fluxes of the form $\Phi=2\pi m /n$ and generic filling. For each $m>1$, we expect the existence of a critical value of $n$, which separates phases without and with overlapping bands. In any case presenting Weyl cones between neighboring bands, TPTs are identified by the topological VH singularities. The behavior of the model for general $m/n$ is non-trivial due to the fractal nature of the spectrum; a natural question, however, would be to verify the existence of a general limit $\Phi_c$ of the critical flux for large $m$ (and study its dependence on the filling). 

Depending on the physical system of interest, the topological VH singularities discussed in this paper can be detected through different methods, e.g. ARPES techniques, scanning tunneling spectroscopy and microscopy, and optical and transport measurements \cite{Lv2021RMP}. The Hofstadter model can be realized in ultracold atom quantum simulations with artificial gauge fluxes \cite{Aidelsburger2013,Miyake2013,Aidelsburger2015,aidelsburger2018}. In this context, several techniques have been successfully applied to detect the presence of band-touching points, including Landau-Zener scattering processes \cite{LihKingPRL2012}, interferometric experiments \cite{DucaScience2015} and Bragg spectroscopy \cite{GotzeNature2010}. Furthermore, very recent works allow for the experimental realization of Weyl semimetals by engineering 2D and 3D spin-orbit couplings \cite{Wang2020}, presenting different methods to locate the position of the Weyl nodes in momentum space and to measure the Berry curvature \cite{CooperRMP2019}.

Finally, we comment that it would be interesting to study the fate of the topological VH singularities and the associated topological phase transitions discussed here in the case of higher-order Weyl metals, such as the ones studied in \cite{HOWSM1_Wang,HOWSM2_Hughes}, and when the dimension $D$ is increased. Moreover, a deserving line of activity could be the study of the role of the topological VH singularities in the determination of the superconducting/superfluid-metal critical temperature transition when an attractive interaction is introduced \cite{Park2020}, especially when it may be tailored in order to preserve the topological properties.

\begin{acknowledgments}
	Discussions with L. Lepori, D. Giuliano, G. Takács and N. Melchioni are gratefully acknowledged. A. T. and P. F. acknowledge hospitality and discussions at the workshop "Low dimensional quantum many body systems" in Heidelberg (12-16 July 2021). M. B. is supported by the Villum Foundation (Research Grant No. 25310).
\end{acknowledgments}

\vfill

\pagebreak
\widetext

\appendix

\section{\label{diagonalization_momentumspace}Diagonalization of the 3D Hofstadter model in momentum space}
The Hofstadter model on a cubic lattice can be solved in momentum space, by taking advantage of the interplay between gauge and translational invariance in the presence of a commensurate background magnetic field. This allows for the introduction of the concept of a magnetic Brillouin zone (MBZ) \cite{Lifschitz-Pitaevski}, which can be defined for every choice of the gauge field $\mathbf{A}(\mathbf{x})$. We will focus on the case of a magnetic field $\Phi=2\pi/n$ in all plaquettes ($n \in \mathbb{N}$), corresponding to a magnetic field ${\bf B}= \Phi \left(1,1,1\right)$. Other fluxes, such as $\Phi=2\pi m/n$ ($m\in\mathbb{N}$), or orientations of the magnetic field, such as $\mathbf{B}\propto(0,1,1)$, yield analogous results.

Due to the presence of the Peierls phases in the Hamiltonian, the cubic lattice can be decomposed in a certain number of independent sublattices, depending on the gauge choice. This arbitrariness can be used to determine the gauge generating the smallest number of the sublattices associated with $\Phi$, i.e. the integer $n$. As analyzed in \cite{Burrello-Trombettoni1}, on a 3D cubic lattice the minimal number of sublattices is given by $n$, and a convenient gauge to work with is the Hasegawa gauge \cite{Hasegawa}, given by
\begin{equation}
\mathbf{A}(\mathbf{x})=\Phi(0,x-y,y-x).
\end{equation}
Within this gauge choice, the MBZ is 
\begin{equation}
\text{MBZ}:\quad k_x\in\bigg[-\frac{\pi}{n},\frac{\pi}{n}\bigg],\;k_y\in\bigg[-\frac{\pi}{n},\frac{\pi}{n}\bigg],\;k_z\in\bigg[-\pi,\pi\bigg],
\label{MBZ_definition}.
\end{equation}
The Hamiltonian can be written in terms of smaller independent blocks, the so-called magnetic bands. For each sublattice we have an associated band, and each one is $n$-fold degenerate. We observe that with $\mathbf{k}\in\text{MBZ}$ the allowed values for the momenta are $N/n^2$, and for each of them the matrix to be diagonalized has size $n\times n$. We then get $N/n$ eigenvalues, each one with degeneracy $n$, matching the dimensionality of the problem in real space, see e.g.  \cite{Burrello-Trombettoni1}.

The expression of the Hofstadter Hamiltonian in momentum space is
\begin{equation}
\mathcal{H}=-t\sum_{\mathbf{k}\in\text{MBZ}}\sum_{\hat{j},s}c^\dag_{s',\mathbf{k}}(T_{\hat{j}})_{s',s}e^{-i\mathbf{k}\cdot\hat{j}}c_{s,\mathbf{k}}+\text{h.c.},
\label{reduced_H_momentumspace}
\end{equation}
where $s$ is an index labeling the sublattices, and the matrices $T_{\hat{j}}$ are
\begin{equation}
T_{\hat{x}}=
\begin{pmatrix}
0 & 1 & 0 & 0\\
0 & 0 & \ddots & 0\\
0 & \cdots & 0 & 1\\
1 & 0 & \cdots & 0
\end{pmatrix},
\qquad
T_{\hat{y}}=e^{-\frac{i\pi}{n}}
\begin{pmatrix}
0 & \cdots & 0 & \varphi_0\\
\varphi_1 & 0 & \cdots & 0\\
0 & \ddots & 0 & 0 \\
0 & 0 & \varphi_{n-1} & 0
\end{pmatrix},
\qquad
T_{\hat{z}}=
\begin{pmatrix}
\varphi_0 & 0 & \cdots & 0\\
0 & \varphi_{n-1} & 0 & 0\\
0 & 0 & \ddots & 0\\
0 & \cdots & 0 & \varphi_1
\end{pmatrix}
\end{equation}
in the sublattice basis. In the previous expressions, we defined for the sake of simplicity $\varphi_l=e^{\frac{2\pi i l}{n}}$, with $l=0,\ldots,n-1$. We observe that for $n\neq2$ the matrices $T_{\hat{j}}$ are not invariant by the conjugate operation, reflecting the fact that time-reversal symmetry is broken by the presence of the external magnetic field. Moreover, the Hamiltonian in Eq. \eqref{hof} has a chiral sublattice symmetry $c_{\mathbf{r}}\rightarrow(-1)^{x+y+z}c_{\mathbf{r}}$ that maps $\mathcal{H}\rightarrow-\mathcal{H}$. As a consequence, the model has a  symmetric single-particle energy spectrum.

\section{\label{general_n_results}Properties of the density of states of the $3D$ Hofstadter model}
\begin{figure}[h]
	\centering
	\includegraphics[width=\linewidth]{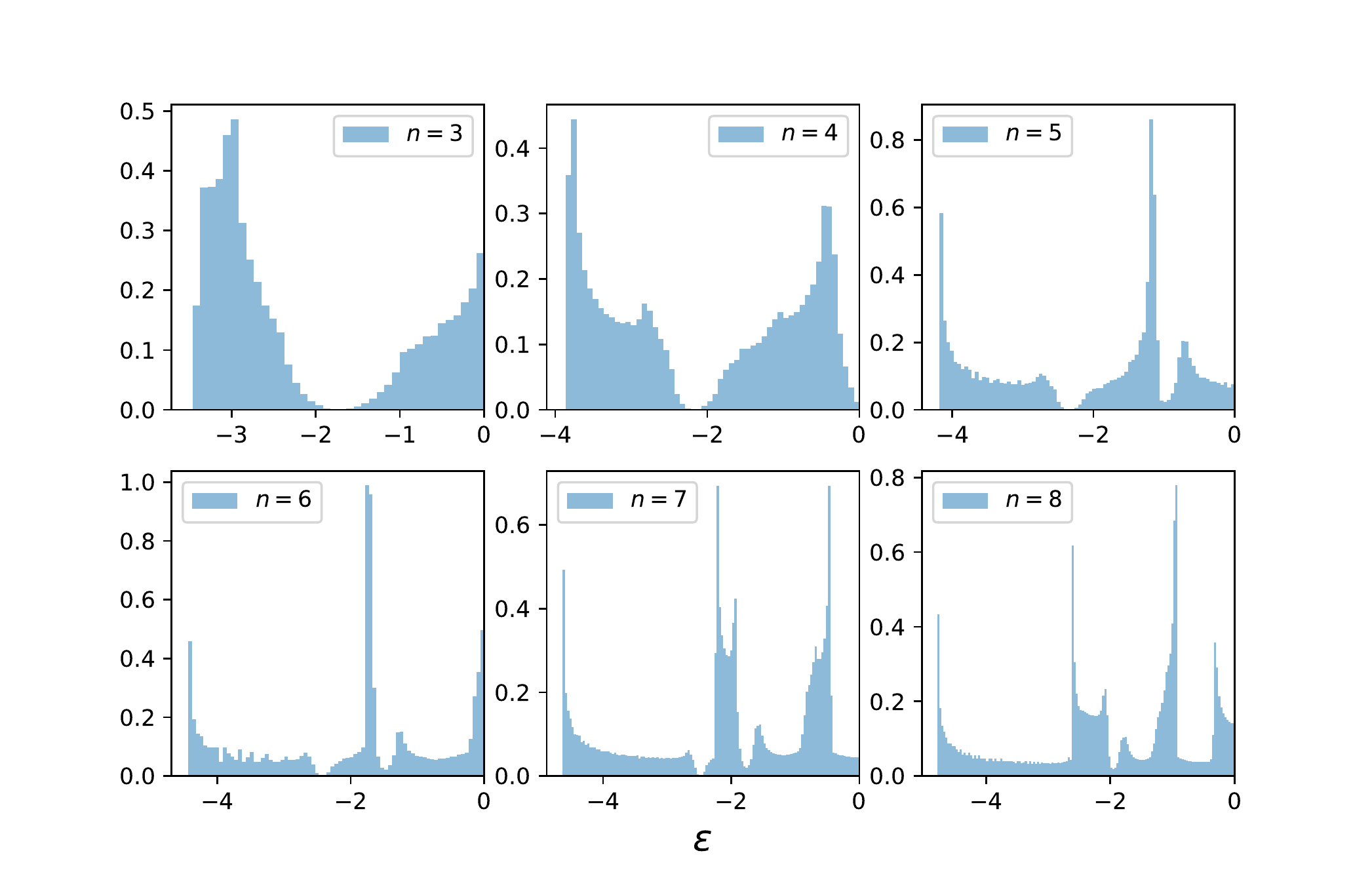}
	\caption{Normalized DOS profiles from $n=3$ to $n=8$, obtained with the exact diagonalization of the Hofstadter Hamiltonian on a finite cubic system of linear size $L=120$ (except for $n=7$, where we considered $L=119$). Because the energy spectrum is symmetric with respect to the origin, we plot the profiles for $\epsilon<0$.}
	\label{DOS_profiles_3n_8n}
\end{figure}

There are some general features that can be identified from the DOS profiles, which are displayed in Fig. \ref{DOS_profiles_3n_8n} for $n \le 8$. We consider, in particular, the behavior of the topological metal phase at filling $\nu \approx 1/n$.

The quantities we adopt to characterize the energy features as a function of $n$ are as follows: the bottom of the lowest band of the system, i.e. the lowest eigenvalue $\epsilon_{min}$ of the Hamiltonian; the energy location $\epsilon_0$ of the first zero of the DOS; and the number $N_w$ of inequivalent Weyl points in the MBZ at $\epsilon_0$. We also numerically checked that the filling $\nu_w$ of the system in the Weyl phase at $\mu=\epsilon_0$ is $1/n$, as expected from the diagonalization of the $n\times n$ Hamiltonian \eqref{reduced_H_momentumspace}. These quantities are summarized in Table \ref{table_various_n} (given the chiral sublattice symmetry, symmetric results hold for $\epsilon>0$).

\begin{figure}[h]
	\centering
	\includegraphics[width=0.32\linewidth]{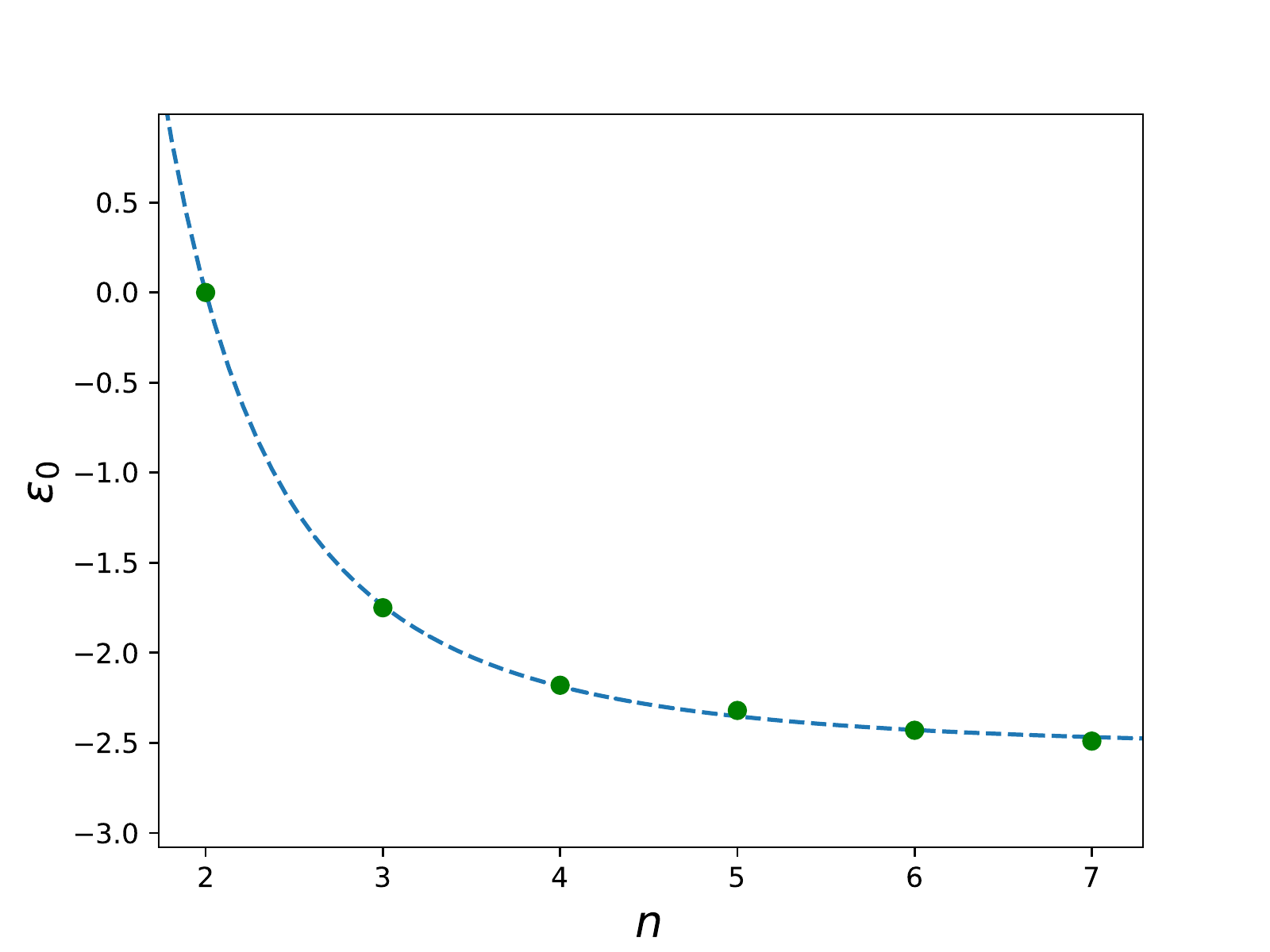}
	\hfill
	\includegraphics[width=0.32\linewidth]{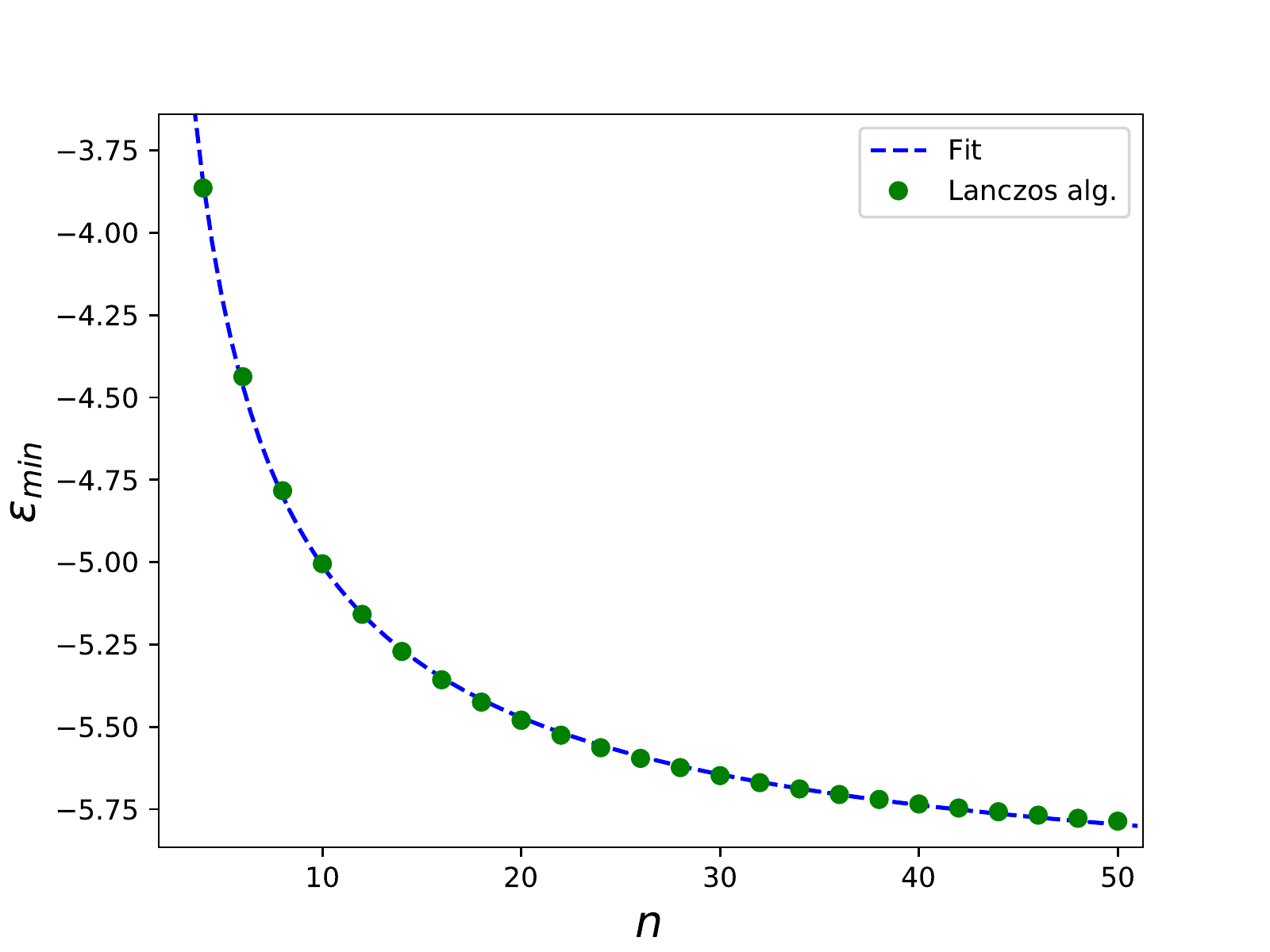}
	\hfill
	\includegraphics[width=0.32\linewidth]{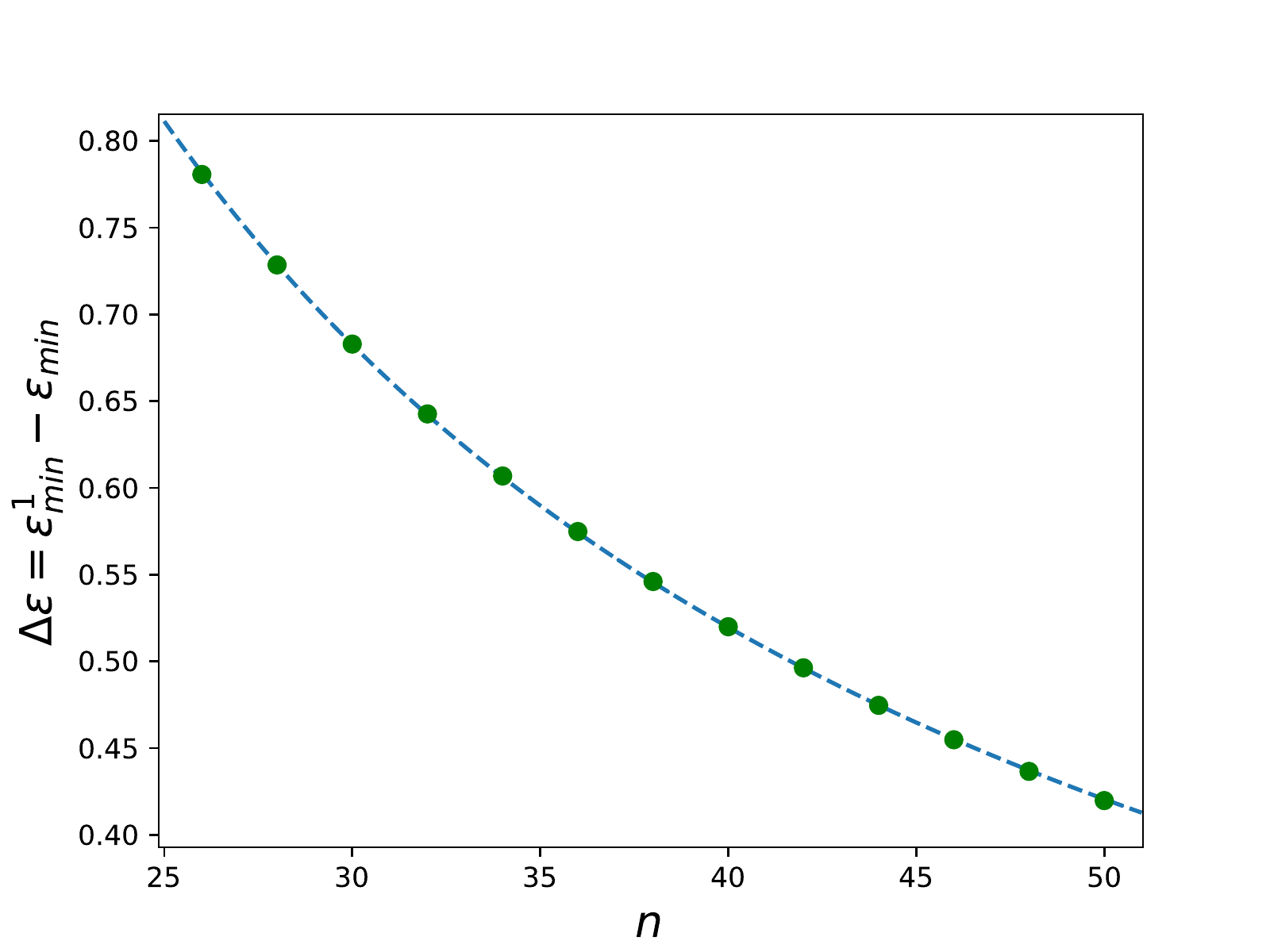}
	\caption{Left panel: Weyl node energy $\epsilon_0$, for $n\leq6$, (a dotted function $\epsilon_0= n^{-2.8} + \kappa $ is depicted as a guide for the eye). Central panel: $\epsilon_{min}$, for $n\leq50$ together with the fit function in Eq. \eqref{fits_n_scaling}. Right panel: $\Delta\epsilon$, for $25 \leq n \leq50$, superimposed with the function $\Delta\epsilon\propto n^{-1}$. All the estimated values are obtained through a finite-size scaling analysis, up to $L=120$.}
	\label{e0_emin_data_fits}
\end{figure}

\begin{table}[h]
	\setlength{\tabcolsep}{11pt}
	\begin{center}
		\begin{tabular}{c|cccc}
			\hline 
			\hline
			$n$ & $\epsilon_{min}$ & $\epsilon_0$ & $N_w$ & $\nu_w$\\  
			\hline
			3 & -3.46 & -1.75 & 2 & 1/3\\
			4 & -3.86 & -2.18 & 2 & 1/4\\
			5 & -4.18 & -2.32 & 2 & 1/5\\
			6 & -4.43 & -2.43 & 2 & 1/6\\
			7 & -4.63 & -2.49 & 2 & 1/7\\
			8 & -4.78 & / & /& / \\
			\hline
			\hline
		\end{tabular}
	\end{center}
	\caption{Parameters as a function of the integer $n$. We report the bottom of the energy bands $\epsilon_{min}$, the energy location $\epsilon_0$, the related number $N_w$ of Weyl points and the filling $\nu_w$ of the Weyl semimetal phase (energies in units of $t$). For $n=8$ the DOS does not display zeros, therefore it is not possible to properly identify $\epsilon_0$, $N_w$ and $\nu_w$ from it. The values are calculated for a discretized MBZ with a mesh of $L^3$ points with $L=120$ (except for $n=7$, where we considered $L=119$). By doing finite-size scaling with a dependence of the form $\alpha+\beta /L^\gamma$, 
		one finds values compatible with those of the table in the first two columns with error $\approx 0.01$.}
	\label{table_various_n}
\end{table}
The position of $\epsilon_0$ is determined through the identification of the DOS zeros. Its value decreases as a function of $n$, and thus for decreasing magnetic fluxes, as can be clearly seen in the left panel of Fig. \ref{e0_emin_data_fits}, where we display the function $\epsilon_0\sim n^{-b}$, with $b\approx-2.8$, as a guide for the eye on top of the estimated values of $\epsilon_0$.

Qualitatively, the same behavior of $\epsilon_0$ is observed for $\epsilon_{\min}$, as shown in the central panel of Fig. \ref{e0_emin_data_fits}. Here we consider magnetic fields up to $n\leq50$, using the Lanczos algorithm \cite{Lanczos} to get the lowest eigenvalue of Eq. \eqref{reduced_H_momentumspace}. To characterize this decreasing behavior, we fit the data with the function
\begin{equation}
\epsilon_{min}=\alpha+\frac{\beta}{n^\gamma}.
\label{fits_n_scaling}
\end{equation} 
The optimal parameters are found to be $\alpha=-6.09(1)$, $\beta=6.8(1)$ and $\gamma=0.79(1)$.

Regarding the number $N_w$ of inequivalent Weyl points in the MBZ, since the system has inversion symmetry but no time-reversal symmetry, its minimum value is $2$, allowing for the existence of a pair of Weyl nodes with opposite momenta at the same energy \cite{Burkov,El-Batanouny}. For $n<8$ we exactly observe two nodes at $\epsilon=\epsilon_0$, as reported in Table \ref{table_various_n}: therefore, within the MBZ, the Hofstadter model hosts the minimum number of nodes compatible with the symmetries of the system, being an example of the so-called ideal Weyl semimetals \cite{SohPRB2019,LuSB2020}. Denoting with $\mathbf{k}_0^{(\pm)}$ the positions of the Weyl nodes in momentum space, the effective Hamiltonian around them can be written as
\begin{equation}
\mathcal{H}(\delta\mathbf{k})=f(\delta\mathbf{k})\;\mathbb{1}+\sum_{i,j=x,y,z}(\delta k)_iv_{ij}\sigma_j,
\label{effective_weyl_generic_n}
\end{equation}
with $\delta\mathbf{k}=\mathbf{k}-\mathbf{k}_0^{(\pm)}$. The function $f(\delta\mathbf{k})$ represents the overall tilt of the Weyl cone, while the tensor $v_{ij}$ contains the velocity vectors. Their specific form can be computed by expanding the Hamiltonian \eqref{reduced_H_momentumspace} around the Weyl points, and then computing the low-energy Hamiltonian \eqref{effective_weyl_generic_n} using degenerate perturbation theory. Since there do not exist directions along which the tilt dominates over the pure Weyl part, the Weyl points are of type I. The contributions of these points to the DOS are still quadratic, but the presence of the tilt and anisotropies makes the computations of the DOS coefficients not analytically doable \cite{Grassano2020EPJB}.

We finally investigate the difference between the minima of the first and the second bands $\Delta\epsilon=\epsilon_{min}^1-\epsilon_{min}$, in order to verify that for $\Phi\ll1$ and filling $\nu\ll1$ the lattice structure becomes negligible, and the system can be described by its continuum counterpart \cite{Hasegawa}. We therefore expect $\Delta\epsilon\sim n^{-1}$, in agreement with a Landau levels description of the model. We plot the energy difference $\Delta\epsilon$ as a function of the magnetic field in Fig. \ref{e0_emin_data_fits}, observing the expected behavior.

\section{\label{HHmodel_VH}Topological van Hove singularities and Weyl points}
\begin{figure}[h]
	\centering
	\includegraphics[width=0.40\linewidth]{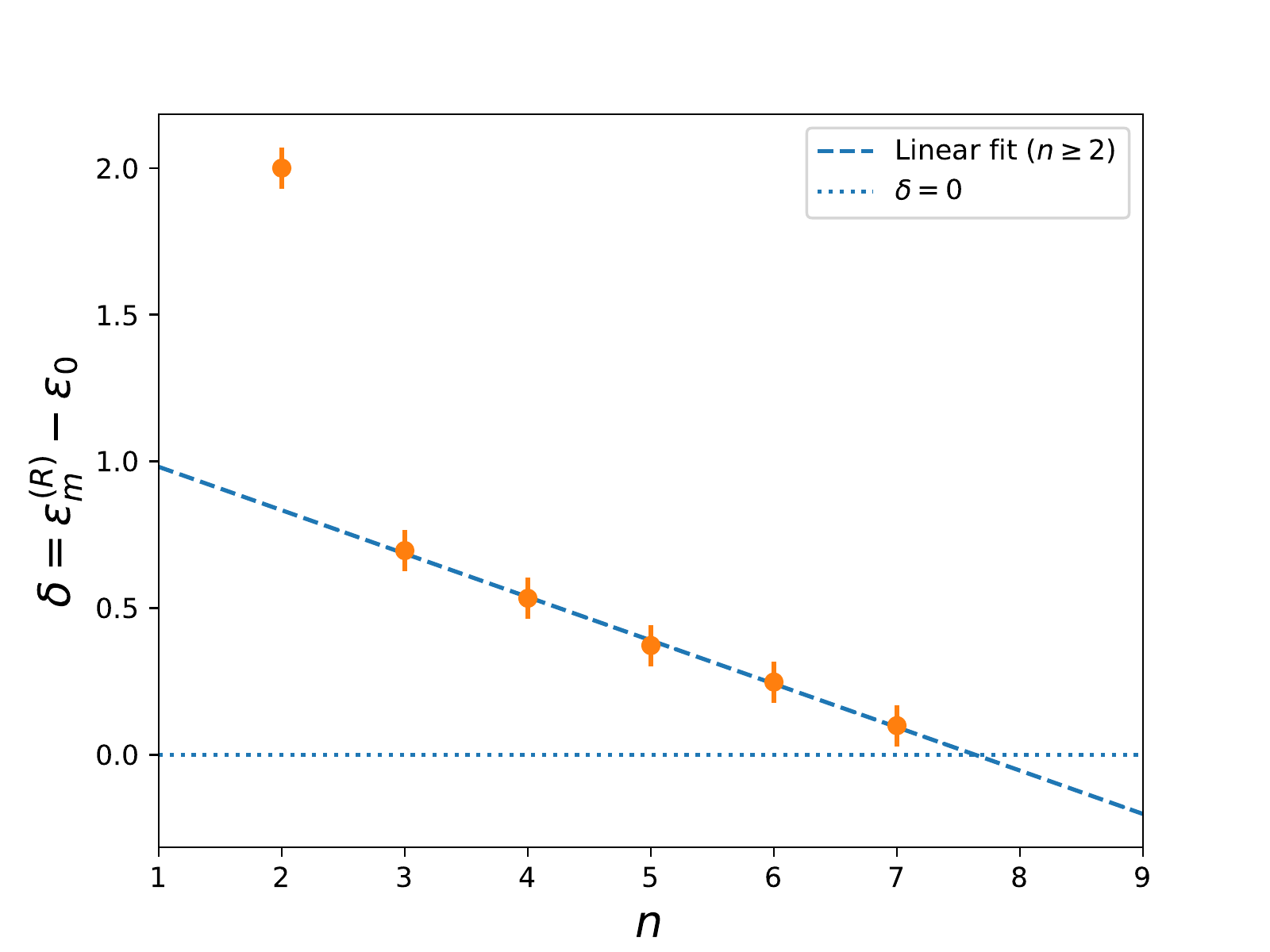}
	\qquad
	\includegraphics[width=0.5\linewidth]{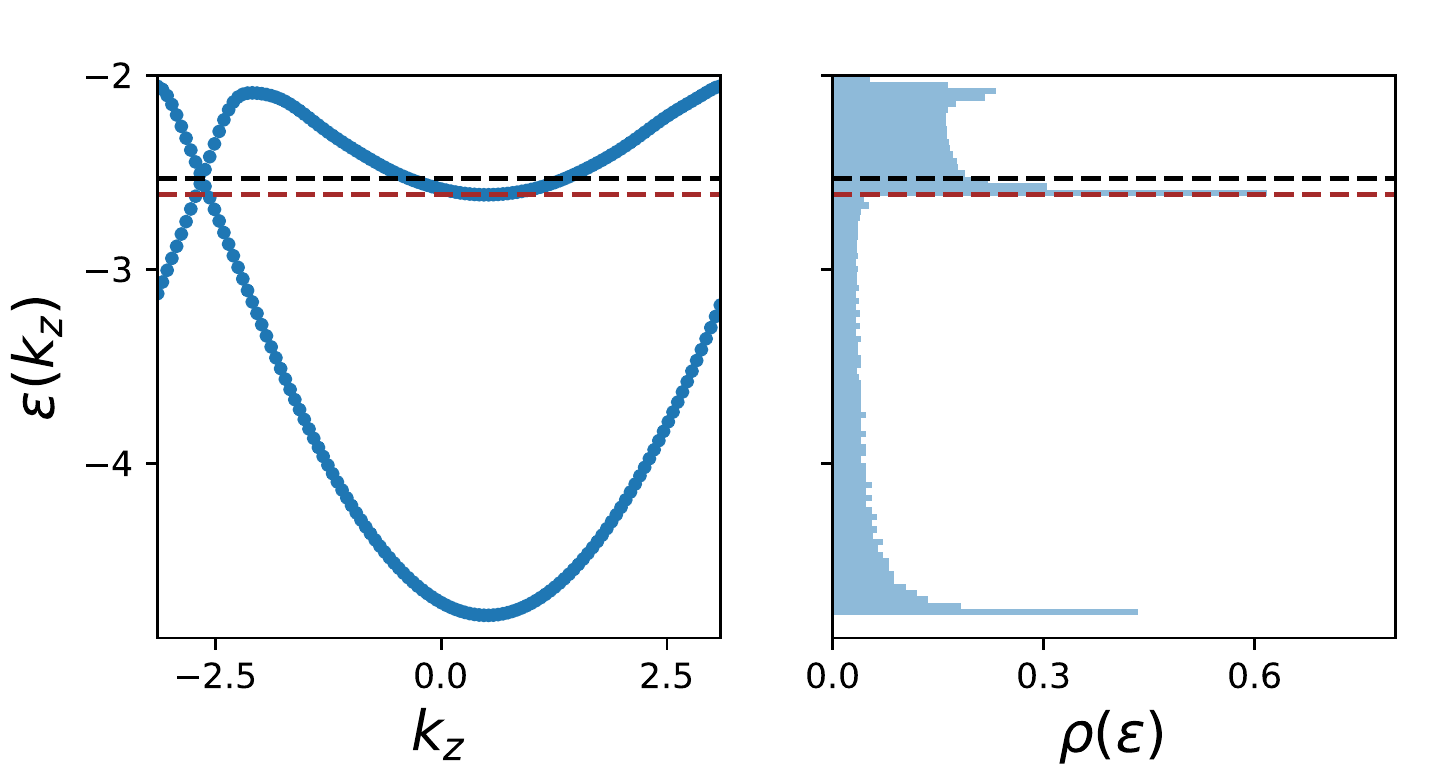}
	\caption{Left plot: energy difference $\delta$ as a function of $n$, for $n\leq8$. We also plot $\delta=\tau n+q$ (see the text), with the estimated parameters values (blue dashed line) and the line $\delta=0$ (black dotted line), for which the energy difference closes. The errorbars in the plot are estimated through a binning analysis on the DOS profiles. Right plot: two lowest energy bands at fixed $k_x=k_x^{(w)}$, $k_y=k_y^{(w)}$ (left panel), alongside the DOS of the model (right panel) for $n=8$. We highlight the VH singularity associated with the minimum of the second band at energy $E_m$ (brown dashed line) and the energy position of the Weyl nodes $E_w>E_m$ (black dashed line).}
	\label{delta_n}
\end{figure}

We investigate in this appendix how the energy difference between $\epsilon_0$ and the topological VH singularity vanished as a function of $n$. Denoting with $\epsilon_m^{(R)}>\epsilon_0$ the energy of the first topological VH singularity found by increasing the energy from the Weyl points, we define the parameter
\begin{equation}
\delta \equiv \epsilon_m^{(R)}-\epsilon_0,
\end{equation}
whose plot as a function of $n$ is shown in Fig. \ref{delta_n}. Apart from the special value of $n=2$, we observe a net linear trend, therefore we interpolate the data with a linear function of the form $\delta=\tau n+q$. The obtained parameters are $\tau=-0.16(1)$, $q=1.20(5)$, and the value of $n$ such that the energy difference closes, i.e. $\delta=0$, corresponds to $7.6(2)$. 

The associated critical value of the flux $\phi_c=2\pi/n_c$ is consistent with the estimate found by Hasegawa \cite{Hasegawa}. In his work, he observed how the energy bands of the Hofstadter model overlap for several pairs $(m,n)$, associated with the flux $\Phi=2\pi m/n$, finding that this happens for fluxes $\phi>\phi_c$, with $\phi_c$ associated with $(m_c,n_c)=(4,31)\sim (1,8)$.

Our analysis on the critical points of the energy bands supports this argument, allowing us to conclude that the contribution of the Weyl point in the DOS for $n\geq8$ is always screened by the presence of other states from the upper band overlapping at the same energy (see, for example, the right panel of Fig. \ref{delta_n}). In this case, the Weyl semimetal phase disappears, and the system enters a more general topological metallic phase. Quantitatively, this is parametrized by the behavior of $\delta(n)$, which linearly closes at $n_c\in(7,8)$. As a consequence, the DOS for $n>n_c$ no longer displays any zero at $\epsilon_0$, despite the presence of Weyl points.

\section{\label{W_tildeWpoints_4n}The case $n=4$ at zero energy}

\begin{figure}[h]
	\centering
	\includegraphics[width=0.4\linewidth]{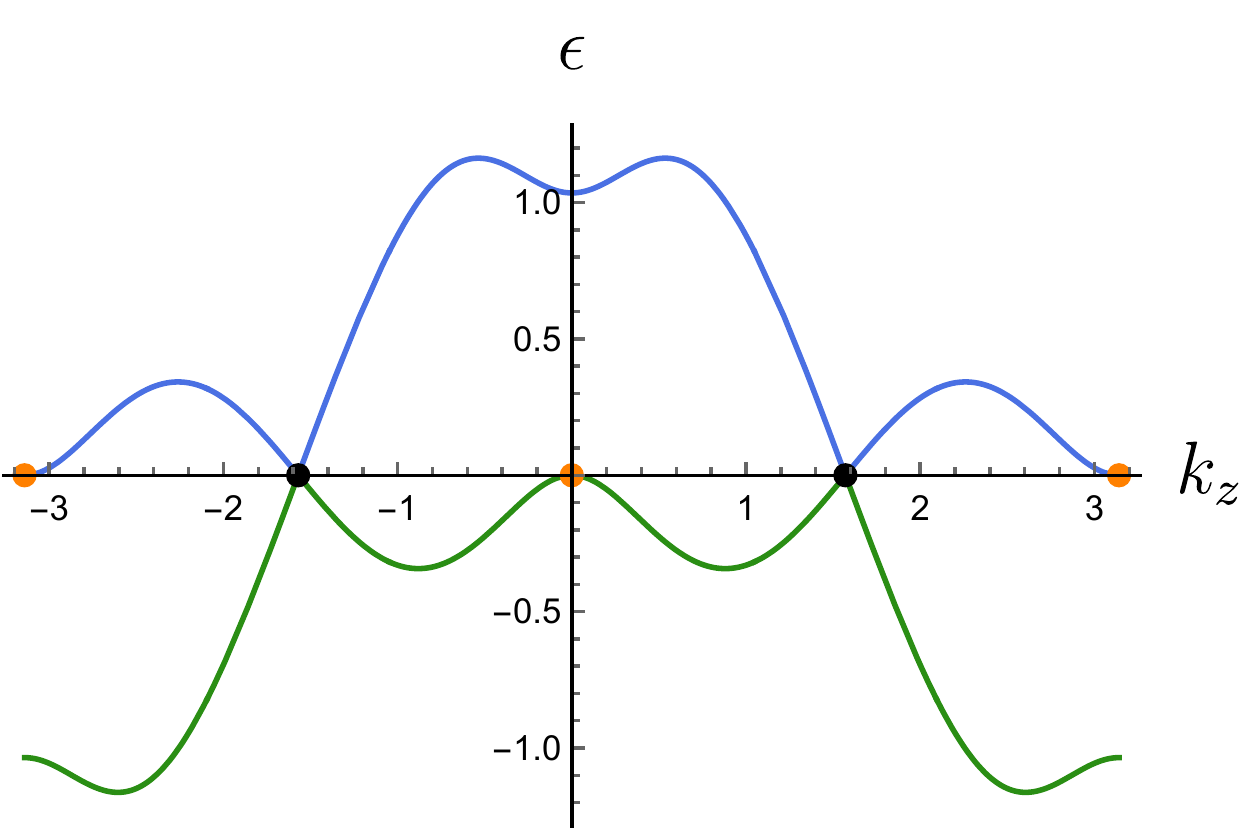}
	\caption{Plot of the central bands of the Hofstadter model with $n=4$ at fixed $k_x=k_y=0$. We highlight the stationary points $\mathbf{k}=(0,0,\pm\pi)$, $(0,0,0)$ at $\epsilon=0$ (orange dots) together with the Weyl points (black dots).}
	\label{4n_centralbands_nodes}
\end{figure}

For $n=4$, the Hofstadter model shows a peculiar behavior at $\epsilon=0$. Indeed, at this energy there are two crossing points for the central bands, exactly in correspondence of the two momenta $\mathbf{k}_0^{(\pm)}=(0,0,\pm\pi/2)$. At the same energy there are also two stationary points of the crossing bands, namely a maximum of the second band at $\mathbf{k}_M=(0,0,0)$ and a minimum of the third band at $\mathbf{k}_m=(0,0,\pi)$, as shown in Fig. \ref{4n_centralbands_nodes}. We expand the dispersion relations around these points, determining the eigenvalues of the Hessian and the behavior of the DOS at $\epsilon=0$. The functional forms of the DOS around them are \cite{Grosso_Parravicini,Bassani_Parravicini}
\begin{equation}
\rho_m(\epsilon\simeq0)=
\begin{cases}
A\sqrt{\epsilon},\qquad\epsilon>0,\\
O(\epsilon),\qquad\epsilon<0,
\end{cases}
\qquad\qquad\qquad
\rho_M(\epsilon\simeq0)=
\begin{cases}
A\sqrt{-\epsilon},\qquad\epsilon<0,\\
O(\epsilon),\qquad\epsilon>0,
\end{cases}
\label{DOS_stationarypoints_4n}
\end{equation}
where the coefficient $A$, which depends on the curvature of the energy band around the stationary point, turns out to be the same for both the functions, since the Hessians of the dispersion relations in $\mathbf{k}_m$ and $\mathbf{k}_M$ are equal and opposite, giving rise to the same modulus of the curvature.

The total contribution to the DOS at $\epsilon=0$ is the sum of the single DOS at the stationary points and at the Weyl points
\begin{equation}
\rho(\epsilon=0)=\rho_m+\rho_M+\rho_W=A\sqrt{|\epsilon|}+B\epsilon^2,
\label{totalDOS_epsilon0}
\end{equation}
where $B$ is a numerical factor encoding the effects of the tilt and the anisotropy of the Weyl cones \cite{Grassano2020EPJB}. From Eq. \eqref{totalDOS_epsilon0} we conclude that the behavior of the DOS around $\epsilon=0$ is not quadratic, due to the presence of stationary points at the same energy of the Weyl ones.

\bibliography{biblio_MB}

\end{document}